\documentclass[aps, prd, superscriptaddress, nofootinbib, showpacs, twocolumn, longbibliography]{revtex4-1}
\usepackage{amsmath, amsfonts, amssymb, mathrsfs}
\usepackage{bm}
\usepackage{graphicx}
\usepackage{hyperref}
\usepackage{color}

\numberwithin{equation}{section}

\newcommand{\tr}{\mathop{\mathrm{tr}}\nolimits}
\newcommand{\Tr}{\mathop{\mathrm{Tr}}\nolimits}

\renewcommand{\Re}{\mathop{\mathrm{Re}}\nolimits}
\DeclareMathOperator*{\Res}{Res}
\newcommand{\calC}{\mathcal C}
\newcommand{\calE}{\mathcal E}
\newcommand{\bbK}{\mathbb K}

\begin {document}

\title{Heat kernel for higher-order differential operators and generalized exponential functions}

\author{A. O. Barvinsky}
\email{barvin@td.lpi.ru}
\affiliation{Theory Department, Lebedev Physics Institute, Leninsky Prospect 53, Moscow 119991, Russia}

\author{P.\,I. Pronin}
\email{petr@phys.msu.ru}
\affiliation{Department of Theoretical Physics, Faculty of Physics, M.\,V. Lomonosov Moscow State University, 119991 Moscow, Russia.}

\author{W. Wachowski}
\email{vladvakh@gmail.com}
\affiliation{Theory Department, Lebedev Physics Institute, Leninsky Prospect 53, Moscow 119991, Russia}

\begin{abstract}
We consider the heat kernel for higher-derivative and nonlocal operators in $d$-dimensional Euclidean space-time and its asymptotic behavior. As a building block for operators of such type, we consider the heat kernel of the minimal operator---generic power of the Laplacian---and show that it is given by the expression essentially different from the conventional exponential WKB ansatz. Rather it is represented by the generalized exponential function (GEF) directly related to what is known in mathematics as the Fox--Wright $\varPsi$-functions and Fox $H$-functions. The structure of its essential singularity in the proper time parameter is different from that of the usual exponential ansatz, which invalidated previous attempts to directly generalize the Schwinger--DeWitt heat kernel technique to higher-derivative operators. In particular, contrary to the conventional exponential decay of the heat kernel in space, we show the oscillatory behavior of GEF for higher-derivative operators. We give several integral representations for the generalized exponential function, find its asymptotics and semiclassical expansion, which turns out to be essentially different for local operators and nonlocal operators of noninteger order. Finally, we briefly discuss further applications of the GEF technique to generic higher-derivative and pseudodifferential operators in curved space-time, which might be critically important for applications of Ho\v{r}ava--Lifshitz and other UV renormalizable quantum gravity models.
\end{abstract}

\maketitle

\section{Introduction}

Physical phenomena in higher derivative and nonlocal field theories are essentially different from conventional local quantum field theory (QFT) with the wave operators of second order in space-time derivatives. There are numerous manifestations of this difference including the problem with unitarity which arises due to higher-derivative (Ostrogradsky) ghosts \cite{Stelle}, violation of Lorentz invariance and peculiar causality properties of Ho\v{r}ava--Lifshitz gravity models (which are motivated by the attempts to solve this problem in renormalizable quantum gravity), noninteger conformal operator dimensions in conformal field theories and so on. These peculiarities are deeply rooted in mathematical formalism of higher derivative models and one of its fundamental ingredients---the heat kernel---the building block underlying the propagator of the theory in Feynman diagrammatic technique.

Now the heat kernel method is one of the most powerful tools in mathematical physics, that has a wide range of applications extending from pure mathematics (spectral geometry) to the analysis of financial markets. Being combined with the background field method in QFT it provides directly in the coordinate space a calculational technique for the quantum effective action, studying renormalizability of field models, their quantum anomalies, critical phenomena, etc.  This makes it indispensable for computations in the presence of external fields or in curved space-time, which is crucially important for gauge theories and quantization of gravity \cite{Gibbons, JackOsborn, JackParker, LeeRim, Barvinsky85, Barvinsky87, CPT2, CPT4, Barvinsky03}. See also \cite{Avram00, Avram01, Vassil03} and references there.

Importance of the heat kernel approach was understood long ago both by mathematicians \cite{Hadamard, Minakshi, Minakshi2, Seeley, Gilkey1975, Gilkey1979} and physicists \cite{Fock, Schwinger,Dewitt}. But the efforts of mathematician were mainly focused on general estimations and theorems regarding the proper time expansion of the \emph{functional trace} of the heat kernel with curvature invariant coefficients \cite{Seeley, Gilkey1975, Gilkey1979, GilkeyGrubb, Gilkey2003, Bar2003, Bogdan2010, Gimperlein2014}, whereas the physicists would consider the two-point heat kernel itself with the separate point arguments \cite{Schwinger, Dewitt, Barvinsky85}. This would give essentially more flexibility and efficiency in obtaining these coefficients---with the ultimate goal of physical applications in UV renormalization and gradient expansion. This is where the difference between the expression for this kernel for second order and higher order operators starts explicitly showing up. Gilkey--Seeley approach, which is based on functorial methods \cite{Gilkey1980, GilkeyFegan, GilkeyBransonFulling, BransonGilkeyPierzchalski}, does not feel this difference, while the Schwinger--DeWitt technique, which explicitly generates recurrent equations for the two-point HaMiDeW (Hadamard--Minakshisundaram--DeWitt) coefficients and their coincidence limits, is very vulnerable to the choice of the leading order heat kernel ansatz and breaks down when it is inappropriately chosen.

Consider a generic {\em minimal} second-order operator $F(\nabla)=-\Delta + \ldots$ whose covariant derivatives form a Laplacian $\Delta = g^{ab}\nabla_a\nabla_b$ acting in a curved $d$-dimensional space-time with the coordinates $\bm{x}=x^a$ and the Riemannian metric $g_{ab}$. Then the ansatz for its heat kernel $\bbK_F(\tau|\bm{x},\bm{y}) = e^{-\tau F(\nabla)}\delta(\bm{x},\bm{y})$ has the form
\begin{multline} \label{HeatKernel0}
\bbK_F(\tau|\bm{x},\bm{y}) = \frac{D^{1/2}(\bm{x},\bm{y})}{(4\pi\tau)^{d/2}} \exp\left(-\frac{\sigma(\bm{x},\bm{y})}{2\tau}\right) \\
\times \sum\limits_{n=0}^\infty \tau^{n}\,a_n[F|\bm{x},\bm{y}],
\end{multline}
where $\sigma(\bm{x},\bm{y}) = l^2(\bm{x},\bm{y})/2$ is the Synge world function and $l(\bm{x},\bm{y})$ is the geodetic distance between the points $\bm{x}$ and $\bm{y}$. In fact this ansatz has a semiclassical nature. This is because its exponential coincides with the principal Hamilton function $S(\tau|\bm{x},\bm{y})=\sigma(\bm{x},\bm{y})/2\tau$  of the particle moving in the $\bm{x}$-space and fictitious imaginary time $\tau$ with the Hamiltonian $F(\nabla)$, and the preexponential factor is just the square root of the Van~Vleck--Morette determinant $D(\bm{x},\bm{y})/(2\tau)^d = \det\left[\partial^2S(\tau|\bm{x},\bm{y})/\partial x^a\partial y^b \right]$. So the expansion in powers of the proper time corresponds to the conventional semiclassical expansion in $\hbar$. Two-point coefficients of this expansion then satisfy simple recurrent equations which can be systematically solved for their coincidence limits $a_n[F|\bm{x},\bm{x}]$---local invariants of the space-time curvature and the coefficients of the operator $F(\nabla)$. Note the property of the expansion \eqref{HeatKernel0} that it isolates essentially singular part of the heat kernel in the exponential, which vanishes in the coincidence limit $\bm{y}=\bm{x}$, whereas important physical information is contained in the HaMiDeW-coefficients of the regular expansion $a_n[F|\bm{x},\bm{x}]$.

It is straightforward to formally extend this semiclassical ansatz to higher-derivative or pseudodifferential operators of the minimal form $F(\nabla) = (-\Delta)^\nu + \ldots$ with some integer or noninteger $\nu$, but this extension fails to generate solvable recurrent relations for the generalized Schwinger--DeWitt coefficients. The origin of this difficulty is that this semiclassical approach fails to perform the resummation of all negative powers of $\tau$ in the exponential, and the infinite power series in $\tau$ turns out to include infinitely many of its negative powers. Essentially singular part of the $\tau$-expansion at $\tau\to 0$ does not get isolated in the exponential and does not seem to vanish in the coincidence limit $\bm{y}=\bm{x}$ as it happens in \eqref{HeatKernel0} for second-order operators.

Apparently due to this difficulty the heat kernel method was only indirectly used in physical applications with higher order operators. Numerous problems like regularization by higher order covariant derivatives or higher derivative theories, namely, $R^2$-gravity \cite{Fradkin1982, AvramBarvinsky}, nonlocal and superrenormalizable models \cite{Tomb, Modesto} and Ho\v{r}ava--Lifshitz models \cite{Barvinsky17, Barvinsky172} were treated by means of the reduction to minimal second-order operators which allow one to use the expansion \eqref{HeatKernel0}.  Such a reduction technique for a wide class of theories with the generalized causality condition was suggested in \cite{Barvinsky85} and actually allowed to circumvent the use of the proper short-time expansion of the form \eqref{HeatKernel0}. Discussion of the heat kernel method for higher-order operators within similar reduction, functorial or other methods can be found in \cite{GilkeyFegan, LeePac, LeePacShin, Pronin1997, Avram97, Avram98}, see also a series of papers by Gusynin \emph{et al.} \cite{Gusynin1989, Gusynin1990, Gusynin1991, GusyninGorbar, GusyninGorbarRomankov, Gorbar}.

Nevertheless, the heat kernels of higher-order differential operators are themselves important as explicit objects, because these kernels represent the building block of Green's functions of these operators, which are needed not only in the UV limit of their coinciding arguments. Moreover, a consistent version of the expansion \eqref{HeatKernel0} for higher-derivative operators and nonlocal operators of pseudodifferential type should be a source of recurrent relations for the generalized HaMiDeW-coefficients, and this generalization is expected to be a much more powerful tool than the reduction technique mentioned above. So possible applications of the standard method to higher-order operators does not make it less interesting to study their heat kernels directly. This is the goal of the present paper.

To understand the nature of the generalization of \eqref{HeatKernel0} for minimal higher-derivative operators it is enough to consider the case of a flat space-time of the Euclidean signature with the world function $\sigma(\bm{x},\bm{y})=(\bm{x}-\bm{y})^2/2$ and the operator $F(\nabla)=(-\Delta)^\nu$---the $\nu$th power of the Laplacian $\Delta = \delta^{ab}\partial_a\partial_b$, so that the standard heat kernel takes the translationally invariant form $e^{\tau\Delta}\delta(\bm{x},\bm{y})=e^{\tau\Delta}\delta(\bm{x}-\bm{y})$ with
    \begin{equation} \label{HeatKernel}
    e^{\tau\Delta}\delta(\bm{x}) = \frac{1}{(4\pi\tau)^{d/2}} \exp\left(-\frac{\bm{x}^2}{4\tau}\right).
    \end{equation}
Then, the basic fact for a generic and not necessarily integer $\nu$ can be formulated as
    \begin{equation} \label{FunkEvol}
    e^{-\tau(-\Delta)^\nu}\delta(\bm{x}) = \frac1{\big(4\pi\tau^{1/\nu}\big)^{d/2}}\, \calE_{\nu, d/2}\Big(-\frac{\bm{x}^2}{4\tau^{1/\nu}}\Big),
    \end{equation}
where $\calE_{\nu, d/2}(z)$ is what we call \emph{generalized exponential function (GEF)}. It is defined as a two-parameter family of functions represented by the Taylor series
    \begin{equation} \label{Em}
    \calE_{\nu,\alpha}(z) = \frac{1}{\nu}\sum\limits_{m=0}^\infty \frac{\Gamma\left(\frac{\alpha+m}{\nu}\right) }{\Gamma(\alpha+m)} \frac{z^m}{m!}.
    \end{equation}
This function obviously reduces to the usual exponential function at $\nu=1$
    \begin{equation}
    \calE_{1,\alpha}(z) = \exp(z)
    \end{equation}
and recovers the standard Gaussian behavior of the heat kernel. On the contrary, for other values of $\nu$ it performs resummation of negative powers of the proper time, which is impossible with the usual semiclassical ansatz.

It should be emphasized that the functions \eqref{FunkEvol} were originally utilized in higher derivative models in \cite{LeePac} and \cite{Gusynin1991}, the latter paper also including their series expansion \eqref{Em}. Later they were used in the context of anomalous diffusion theory  \cite{ZUS} and in application to Ho\v{r}ava--Lifshitz models \cite{Mamiya14}. However, thus far no systematic studies of these functions were undertaken and their potential role for the extension of the HaMiDeW technique to modified field and gravity models was underestimated. The goal of this work is to fill up this omission.

The paper is organized as follows. In Sec.~\ref{S1} we derive the heat kernel of the operator $(-\Delta)^\nu$ and its associated GEF in the form of the Taylor expansion and present its integral representation in terms of Bessel functions. In Sec.~\ref{S2} we discuss the properties of the generalized exponential functions $\calE_{\nu, \alpha}(z)$ and consider their  Mellin--Barnes integral representation generating their asymptotic behavior at $z\to\infty$, which is responsible for the short time, $\tau\to 0$, or large $|\bm{x}|\to\infty$ limit of the heat kernel \eqref{FunkEvol}. Interestingly, this limiting behavior turns out to be different for fractional and integer powers $\nu$. Contrary to the second-order case this asymptotics is power-law for fractional $\nu$ and exponential for integer $\nu$, and moreover features oscillations for growing $|\bm{x}|$. For integer powers $\nu$ this property is demonstrated in Sec.~\ref{S3} where the asymptotic behavior of GEF is compared with the semiclassical heat kernel ansatz and the saddle-point approximation for the momentum space integral representation. In concluding section we briefly discuss further application of GEF to generic minimal and nonminimal higher-derivative operators in curved space-time, which will allow us to build a solvable recurrent equations for the full set of generalized HaMiDeW-coefficients.

\section{The heat kernel of the power of Laplacian} \label{S1}
For the operator $F=(-\Delta)^\nu$ its heat kernel
    \begin{equation}
    \bbK_{\nu,d}(\tau, \bm{x}) = e^{-\tau(-\Delta)^\nu}\delta(\bm{x})
    \end{equation}
has an obvious momentum space representation
    \begin{equation} \label{UIntP}
    \bbK_{\nu,d}(\tau,\bm{x}) = \int \frac{d^d\bm{k}}{(2\pi)^d} \exp\left(-k^{2\nu}\tau + i\bm{k}\bm{x} \right),
    \end{equation}
where $k=|\,\bm{k}\,|=\sqrt{\bm{k}^2}$ and $\bm{k}\bm{x}=k_ax^a$. For $\nu=1$ this integral defines the well-known fundamental solution \eqref{HeatKernel}.

Note that the heat kernel \eqref{UIntP} is invariant with respect to $O(d)$-rotations and homogeneous
    \begin{gather}
    \bbK_{\nu, d}(\tau,\bm{x}) = \bbK_{\nu, d}(\tau,|\,\bm{x}\,|), \\
    \bbK_{\nu, d}(c^{2\nu}\tau, c\bm{x}) = c^{-d} \bbK_{\nu, d}(\tau,\bm{x}),
    \end{gather}
where $c$ is an arbitrary constant. Therefore, it should have the form \eqref{FunkEvol}, where $\calE_{\nu, d/2}(z)$ is some unknown function of the ratio $-\bm{x}^2/4\tau^{1/\nu}$. Since it stands in place of the exponential function in the usual expression for the heat kernel \eqref{HeatKernel}, we will call it \emph{the generalized exponential function (GEF)}.

Let us find the expansion of the generalized exponential function $\calE_{\nu, d/2}(z)$ in its Taylor series. Using the relations
\begin{equation}
\partial_a \sigma^k = k\sigma^{k-1}x_a, \qquad \partial_a\big(\sigma^k x^a\big) = (d + 2k)\sigma^k,
\end{equation}
where $\sigma = \bm{x}^2/2$, it is easy to verify by induction that for an arbitrary function $f(c\sigma)$, where $c$ is a constant, the following differentiation rule holds
    \begin{equation}
    \Delta^m f(c\sigma) = (2c)^m \sum\limits_{k=0}^m C_m^k \frac{\Gamma(d/2+m)}{\Gamma(d/2+k)}(c\sigma)^k f^{(m+k)}(c\sigma),
    \end{equation}
where $f^{(k)}(z) = d^kf(z)/dz^k$, $C_m^k = m!/k!(m-k)!$ are the binomial coefficients and $\Gamma(s)$ is the Euler gamma function. Then for $f(z) = \calE_{\nu,d/2}(z)$ and $c = -1/2\tau^{1/\nu}$ we obtain at $\bm{x}=0$
\begin{equation} \label{Proizv2}
(-\Delta)^m \bbK_{\nu,d}(\tau,\bm{x})\,\Big|_{\,\bm{x}=0} = \frac{\tau^{-\frac{d/2+m}{\nu}}}{(4\pi)^{d/2}} \frac{\Gamma(d/2+m)}{\Gamma(d/2)} \calE_{\nu,d/2}^{(m)}(0).
\end{equation}
On the other hand, these quantities can easily be calculated directly
\begin{multline} \label{Proizv1}
(-\Delta)^m \bbK_{\nu,d}(\tau,\bm{x})\,\Big|_{\,\bm{x}=0} = \int \frac{d^d\bm{k}}{(2\pi)^d} k^{2m} \exp \left(- k^{2\nu}\tau \right) \\
= \frac{\tau^{-\frac{d/2+m}{\nu}}}{(4\pi)^{d/2}} \frac{\Gamma\left(\frac{d/2+m}{\nu}\right)}{\nu\Gamma(d/2)}.
\end{multline}
Comparing the expressions \eqref{Proizv2} and \eqref{Proizv1}, we find all the derivatives of $\calE_{\nu, \alpha}(z)$ at $z=0$ and thus get its Taylor expansion \eqref{Em}.

\subsection{Bessel functions representation} \label{S12}
Another integral representation expresses the generalized exponential function $\calE_{\nu,\alpha}(z)$ in terms of the Bessel function $J_\alpha(z)$ or the Bessel-Clifford function $\calC_\alpha(z)$. The latter is determined by the series
    \begin{equation} \label{BessCliff}
    \calC_{\alpha}(z) = \sum\limits_{m=0}^\infty \frac{1}{\Gamma(\alpha+1+m)}\frac{z^m}{m!}
    \end{equation}
and related to the Bessel function $J_\alpha(z)$ by the equation which removes its branching point at $z=0$
    \begin{equation} \label{Bessel}
    J_\alpha(z) = \left(\frac{z}{2}\right)^\alpha \calC_\alpha(-z^2/4).
    \end{equation}
Therefore, the Bessel--Clifford functions have no singularities and represent single-valued entire functions on the whole complex $z$ plane.

Bessel function representation of $\bbK_{\nu,d}(\tau,\bm{x})$ follows from integration over angles in the momentum space integral \eqref{UIntP}, which reads as
\begin{multline} \label{IntUgol}
\bbK_{\nu,d}(\tau;\bm{x}) = \frac{S_{d-2}}{(2\pi)^d} \int\limits_0^\infty dk\, k^{d-1} \exp(-k^{2\nu}\tau) \\
\times \int\limits_0^\pi d\theta\,(\sin\theta)^{d-2}\, e^{ikx\cos\theta},
\end{multline}
where $S_{d-2} = 2\pi^{(d-1)/2}/\Gamma(\frac{d-1}2)$ is the volume of $(d-2)$-dimensional unit sphere, $\theta$ is the angle between the vectors $\bm{x}$ and $\bm{k}$ in \eqref{UIntP} and $x=|\,\bm{x}\,|$. Expanding $\exp(ikx\cos\theta)$ and integrating over $\theta$ on account of
\begin{gather}
2\int\limits_0^{\pi/2} (\sin\theta)^{2\alpha-1} (\cos\theta)^{2\beta-1} d\theta = \frac{\Gamma(\alpha)\Gamma(\beta)}{\Gamma(\alpha+\beta)}, \\
\sqrt{\pi}\Gamma(2z+1) = 4^z\, \Gamma\left(z+\frac{1}{2}\right) \Gamma(z+1) \label{Legendre}
\end{gather}
one gets
\begin{multline} \label{ThetaInt}
\int\limits_0^\pi d\theta\,(\sin\theta)^{d-2}\, e^{ikx\cos\theta} \\
= \sqrt{\pi}\Gamma\left(\frac{d-1}{2}\right) \calC_{\frac{d}{2}-1}\left(-k^2x^2/4\right).
    \end{multline}
As a result the heat kernel has the following integral representation with the Bessel-Clifford function
\begin{multline} \label{k2Int}
\bbK_{\nu,d}(\tau,\bm{x}) = \frac{2}{(4\pi)^{d/2}} \int\limits_0^\infty k^{d-1} \exp(-k^{2\nu}\tau) \\
\times \calC_{\frac{d}{2}-1}\left(-k^2x^2/4\right) dk.
\end{multline}
Under the change of integration variable $\mu=k^{2\nu}\tau$ comparison of this expression with \eqref{FunkEvol} gives the relevant integral representations for the GEF $\calE_{\nu,\alpha}(z)$
    \begin{equation} \label{calEIntRep}
    \calE_{\nu,\alpha}(z) = \frac{1}{\nu} \int\limits_0^\infty d\mu\, \mu^{\alpha/\nu-1} e^{-\mu}\, \calC_{\alpha-1}(z\mu^{1/\nu}).
    \end{equation}
Note that substitution of the expansion \eqref{BessCliff} for the Bessel-Clifford function into \eqref{calEIntRep} directly leads to the expansion \eqref{Em}, which confirms this representation.

Another remark is that the functions $\calE_{\nu, \alpha}(z)$ and $\bbK_{\nu,d}(\tau;\bm{x})$ are directly related to Bessel-Clifford \eqref{BessCliff} and Bessel functions \eqref{Bessel} in the limit $\nu\to\infty$. Indeed, replacing in this limit $\Gamma\big((\alpha+m)/\nu\big)$ by $\nu/(\alpha+m)$ in the expansion \eqref{Em}, one has
    \begin{gather}
    \calE_{\infty,\alpha}(z) = \calC_\alpha(z), \label{NuInfty} \\
    \bbK_{\infty,d}(\tau,\bm{x}) = \frac{1}{(2\pi x)^{d/2}} J_{d/2}(x). \label{UInfty}
    \end{gather}
Interestingly, the heat kernel of $(-\Delta)^\infty$ becomes independent of the proper time parameter $\tau$ because of the obvious limit $\tau^{1/\nu}\xrightarrow[\nu\to\infty]{} 1$.

\section{Generalized exponential functions and their properties} \label{S2}
Various properties of GEF follow from the Mellin-Barnes integral representation of this function. This representation can be obtained by converting the series \eqref{Em} into the contour integral in the complex plane of an auxiliary parameter $s$, such that the residues at simple poles of the integrand generate various terms of this series. It is easy to guess that this integral reads
    \begin{equation}
    \calE_{\nu,\alpha}(-z) = \frac{1}{2\pi i} \int\limits_C ds\, \frac{\Gamma(s)\Gamma\left(\frac{\alpha-s}{\nu}\right)}{\nu\Gamma(\alpha-s)}\, z^{-s}. \label{MellinBurns}
    \end{equation}
Then the inverse Mellin transform obviously gives
    \begin{equation}
     \int\limits_0^\infty dz\,z^{s-1} \calE_{\nu,\alpha}(-z) = \frac{\Gamma(s)\Gamma\left(\frac{\alpha-s}{\nu}\right)}{\nu\Gamma(\alpha-s)}
     \equiv \varepsilon_{\nu,\alpha}(s). \label{MellinCalE}
    \end{equation}

\begin{figure}
\includegraphics[scale=2]{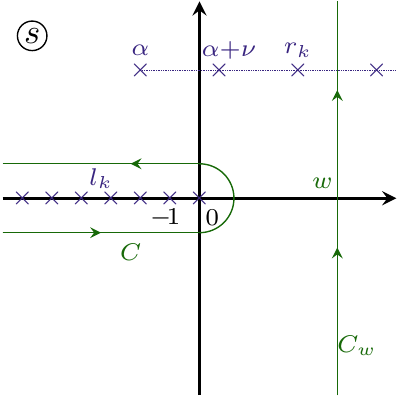}
\caption{\label{Fig4} The location of the poles of $\varepsilon_{\nu,\alpha}(s)$ and the contours $C$ and $C_w$ on the complex $s$ plane.}
\end{figure}

The location of the poles of $\varepsilon_{\nu,\alpha}(s)$ and the contour $C$ on the complex $s$ plane is schematically shown in Fig. \ref{Fig4}. The function $\Gamma(s)$ has a sequence of poles at $l_k = -k$, $k=0,1,2,\ldots$ running to the left with residues $(-1)^k/k!$. And the function $\Gamma((\alpha-s)/\nu)$ has poles at $r_k = \alpha+k\nu$, running to the right. The contour $C$ begins at $-\infty-i\epsilon$, runs under the real axis, bends around 0 and returns to $-\infty+i\epsilon$. The integral \eqref{MellinBurns} is equal to the sum of residues at the poles $l_k$, which exactly gives the series \eqref{Em}.

It is possible, however, that not all the points $r_k = \alpha+k\nu$ are poles of the function $\varepsilon_{\nu,\alpha}(s)$, since they can be canceled by the poles of the function $\Gamma(\alpha-s)$ in the denominator. In particular, the point $r_0 = \alpha$ is never a pole of $\varepsilon_{\nu,\alpha}(s)$. Three cases are possible: if $\nu$ is irrational, then all other points $r_k$ are poles; if $\nu = p/q$ is an irreducible fraction, then the poles $r_{qk}$ are also canceled; and, finally, if $\nu$ is integer, then all the poles $r_k$ are canceled.

To determine the asymptotic behavior of the function $\calE_{\nu,\alpha}(-z)$, we deform the integration contour $C$ into the contour $C_w$, coming from $w-i\infty$ vertically to $w+i\infty$, where $w\ne\Re r_k$, $w>0$ (see Fig. \ref{Fig4}). Then we have
\begin{multline} \label{Intw}
\calE_{\nu,\alpha}(-z) = - \sum\limits_{\Re r_k < w} \Res\limits_{s = r_k}\left[\,\varepsilon_{\nu,\alpha}(s)z^{-s}\right] \\
+ \frac{1}{2\pi i} \int\limits_{w-i\infty}^{w+i\infty} ds\,  \varepsilon_{\nu,\alpha}(s) z^{-s}.
\end{multline}

For $z\to\infty$ the residues decrease as powers $z^{-r_k}$, whereas the integral term is obviously $O(z^{-w})$ because of the constant factor $z^{-w}$ in the integrand. Then for noninteger $\nu$ by pushing $w\to+\infty$ we obtain the sum of all residues as the power-like asymptotic expansion of GEF
    \begin{equation} \label{Asympt}
    \calE_{\nu,\alpha}(-z)= -z^{-\alpha} \sum\limits_{m=1}^\infty \frac{\Gamma(\alpha+m\nu)}{\Gamma(-m\nu)} \frac{(-z^{-\nu})^m}{m!}
    +O(z^{-\infty}).
    \end{equation}

Note that the cancellation of the pole in Mellin image \eqref{MellinCalE} results in the vanishing of the corresponding term in this expansion \eqref{Asympt}. In the case of integer $\nu$ all residue terms vanish. This means that in this case the function $\calE_{\nu, \alpha}(-z)$ decreases faster than any power of $z$, i.e. in an exponential manner. Then the function $\calE_{\nu,\alpha}(-z)$ is exactly equal to the integral over the contour $C_w$ for any positive $w$, and its exponential asymptotic behavior is entirely determined by this integral. This case is very special, and we will consider it in detail later in Sec.~\ref{S3}.

\begin{figure}
\includegraphics[scale=0.8]{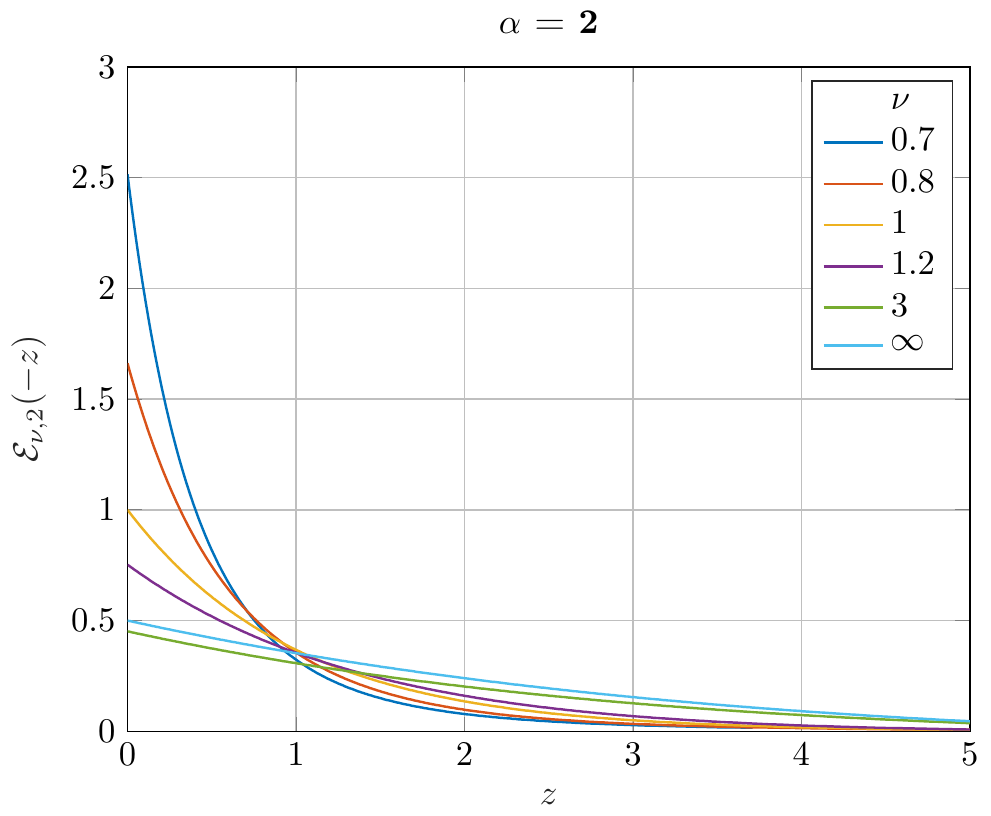}
\includegraphics[scale=0.8]{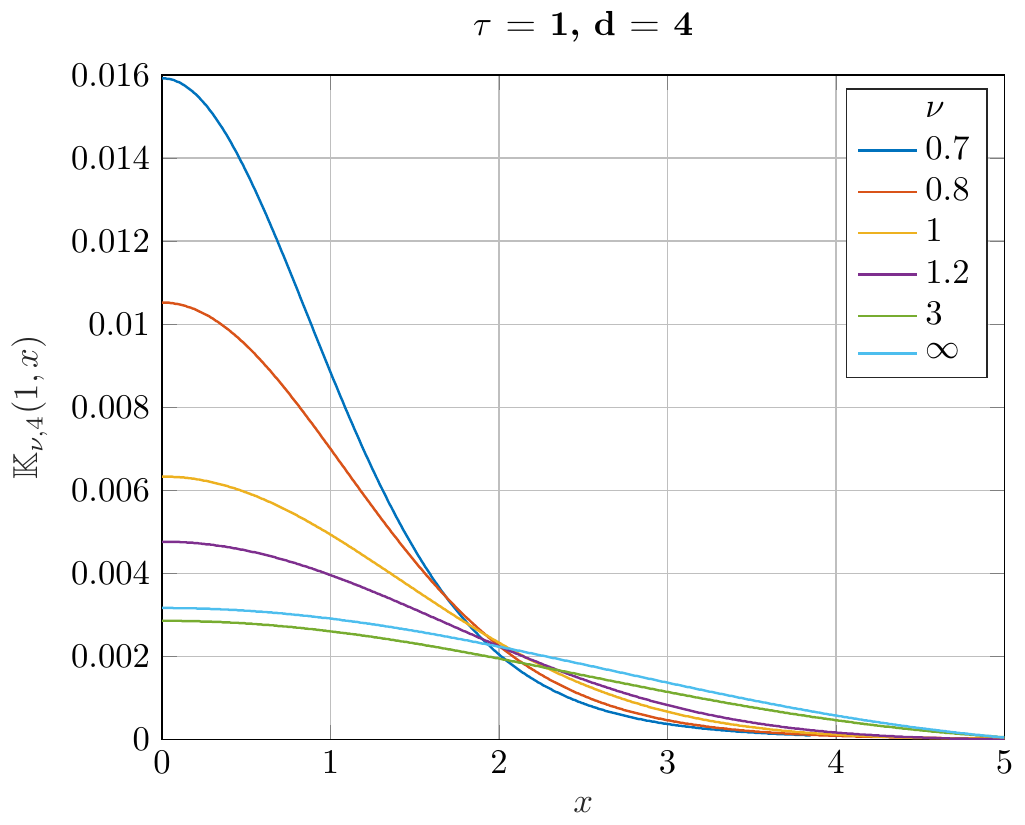}
\caption{\label{Fig1} Graphs of the functions $\calE_{\nu, 2}(-z)$ and $\bbK_{\nu,4}(1,x)$ for various values of the parameter $\nu$. For $\nu = \infty$ the functions are given by \eqref{NuInfty} and \eqref{UInfty}.}
\end{figure}

The properties of GEF of the above type are strongly associated with the properties of ratios of gamma functions and their products, which are in fact a part of the theory of the Fox-Wright $\varPsi$-functions \cite{Wright35, Wright40} and more general Fox $H$-functions \cite{Braaksma, Marichev, Sri84, MSH, KS}. Namely, our generalized exponential function, defined by the series \eqref{Em}, is a special case of the Fox-Wright function ${}_p\Psi_q[\,(a,A); (b,B); z\,]$ labeled by numerous parameters
    \begin{equation} \label{EPsi}
    \calE_{\nu,\alpha}(z) = \frac{1}{\nu} {}_1\Psi_1\big[\, \left(\alpha/\nu,1/\nu\right); (\alpha, 1); z\, \big].
    \end{equation}
For completeness we briefly give the definition of these functions in Appendix \ref{App1}, in particular because their theory gives subtle details of the domain of parameters $\nu$ and $\alpha$, where the function $\calE_{\nu,\alpha}(z)$ is well defined.

For the generalized exponential function $\calE_{\nu,\alpha}(z)$ we have the Mellin--Barnes integral \eqref{MellinBurns} and two power series: the first series \eqref{Em} in $z$ near 0 and the second series \eqref{Asympt} in $1/z$ near $\infty$. As it follows from the general theory of $H$-functions (see Eqs.~\eqref{mu_beta}, \eqref{a} and the statement below them), there are three possible cases: for $\nu>1/2$ the first series converges absolutely on the whole complex plane $z$, and $\calE_{\nu,\alpha}(z)$ is thus an entire function and has the essential singularity at $z=\infty$, while the second series diverges and is asymptotic for $z\to\infty$. When $\nu<1/2$ the situation is the opposite: the series \eqref{Asympt} absolutely converges on the whole complex plane $z$, except for the essential singularity at $z=0$, and the series \eqref{Em} diverges and is asymptotic for $z\to0$. Finally, for the critical value $\nu=1/2$ the series \eqref{Em} converges inside the circle $|z|<1/4$ and diverges outside it, while conversely the series \eqref{Asympt} converges outside this circle and diverges inside it.

It turns out that for the critical value $\nu=1/2$ the series \eqref{Em} can be summed up analytically\footnote{We are grateful to the anonymous referee for pointing out this fact.}. Indeed, using the Legendre duplication formula \eqref{Legendre} and the well-known expansion
\begin{equation}
(1-z)^{-\gamma} = \frac{1}{\Gamma(\gamma)} \sum\limits_{m=0}^\infty \Gamma(\gamma+m) \frac{z^m}{m!},
\end{equation}
we can easily find that
\begin{multline} \label{OneHalf}
\calE_{\frac{1}{2},\alpha}(z) = \frac{4^\alpha}{\sqrt{\pi}} \sum\limits_{m=0}^\infty \Gamma\left(\alpha+\frac{1}{2}+m\right) \frac{(4z)^m}{m!} \\
= \frac{4^\alpha\, \Gamma\left(\alpha+\frac{1}{2}\right)}{\sqrt{\pi}} (1-4z)^{-\alpha-\frac{1}{2}}.
\end{multline}
Thus, for $\nu=1/2$ GEF not only have power-law asymptotic behavior, but they really are power functions. The series \eqref{Em} diverges for $|z|>1/4$ due to the existence of a pole at the point $z=1/4$. It is not difficult to verify that in this case even terms of the series \eqref{Asympt} vanish and odd terms converge to the function \eqref{OneHalf} in the circle $|z|>1/4$. As a result, the operator $\sqrt{-\Delta}$ in a flat $d$-dimensional space has the following heat kernel
\begin{equation} \label{OneHalfKern}
\bbK_{\frac{1}{2},d}(\tau, \bm{x}) = \frac{\Gamma\left(\frac{d+1}{2}\right)}{\pi^\frac{d+1}{2}} \frac{\tau}{\left(\tau^2+\bm{x}^2\right)^\frac{d+1}{2}}.
\end{equation}
This expression is related to holographic \cite{Witten, LiuTseytlin} and brane world \cite{LPR, BarvinskyNesterov06, BarvinskyNesterov10} applications of effective action because it represents the massless limit of the simplest brane-to-bulk propagator $e^{-\tau\sqrt{M^2-\Box}}\delta(\bm{x})$, $M\to 0$, \cite{BarvinskyNesterov06, BarvinskyNesterov10} and may be interesting in the context of the discussion of fractional powers of generalized Laplacians in \cite{Bar2003}.

The terms of the series \eqref{Em} for $\calE_{\nu, \alpha}(z)$ are well defined for complex
    \begin{equation} \label{UslOpr}
    \alpha\ne-n-k\nu, \quad\text{where}\quad n,k = 0,1,2,\ldots
    \end{equation}
We note, however, that the singularities at the points $\alpha = -n$ (and hence at all the points for integers $\nu$) are removable, since the poles of the gamma functions in the numerator and denominator cancel each other. Expanding $\Gamma(-n+\epsilon) \sim (-1)^n/n!\epsilon$ we can redefine the coefficients in \eqref{Em}
\begin{equation}
\frac{\Gamma(-n/\nu)}{\nu\Gamma(-n)} = \begin{cases}
(-1)^{-n+n/\nu}\frac{n!}{(n/\nu)!}, & \text{if }n/\nu = 0,1,\ldots, \\
0, & \text{otherwise}. \end{cases}
\end{equation}

Thus, for integer $\nu$ the function $\calE_{\nu, \alpha}(z)$ is an entire function of $z$ for any values of $\alpha$, and for noninteger $\nu>1/2$ it is an entire function for all values of $\alpha$ except $\alpha= -n-k\nu\neq -m$ with positive integer $k,n$ and $m$. Consequently, the functions $\bbK_{\nu, d}(\tau, \bm{x})$ are well defined not only for all integer $\nu$ and $d$, but also for fractional $\nu$ and $d$ satisfying these conditions.

\begin{figure}
\includegraphics[scale=0.8]{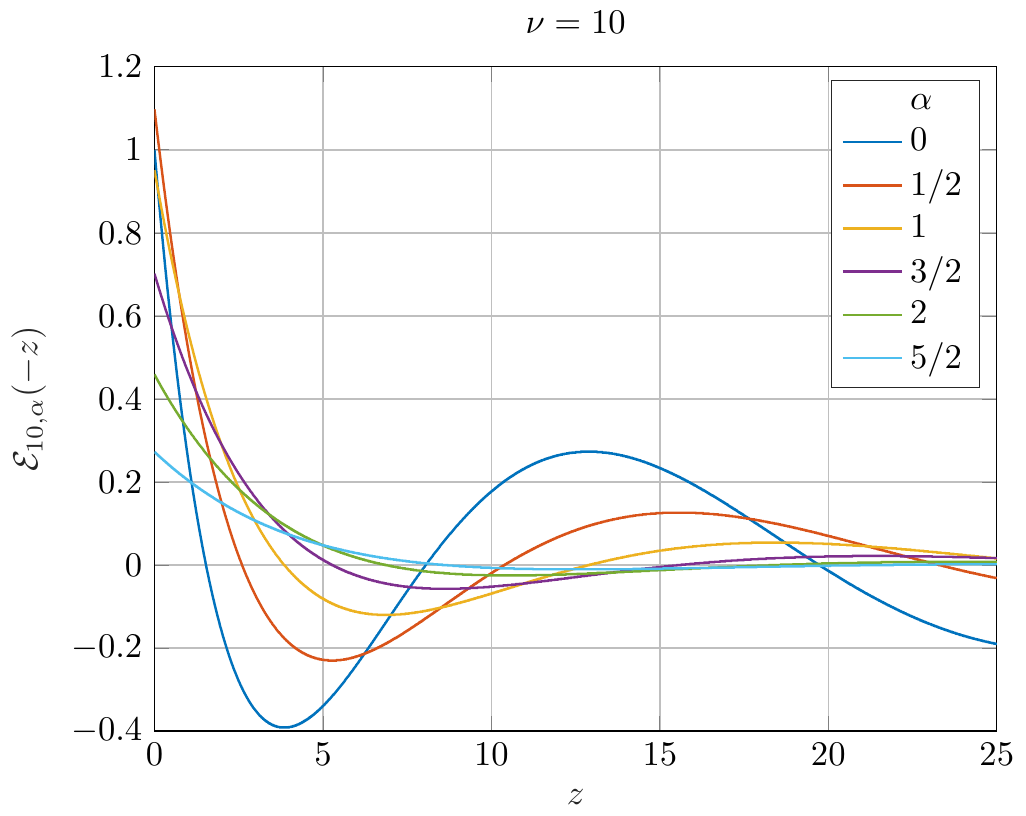}
\includegraphics[scale=0.8]{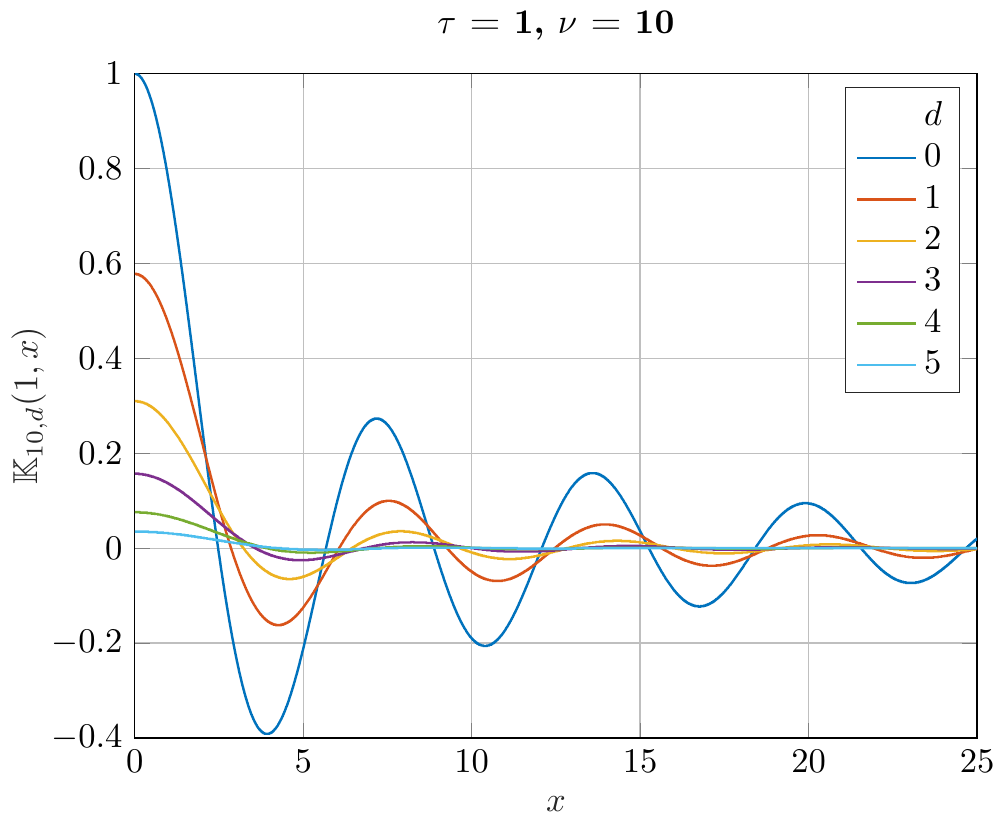}
\caption{\label{Fig2} Graphs of the functions $\calE_{10,\alpha}(-z)$ and $\bbK_{10,d}(1, x)$ for various values of the parameters $\alpha$ and $d$.}
\end{figure}

\begin{figure} \begin{center}
	\includegraphics[scale=0.8]{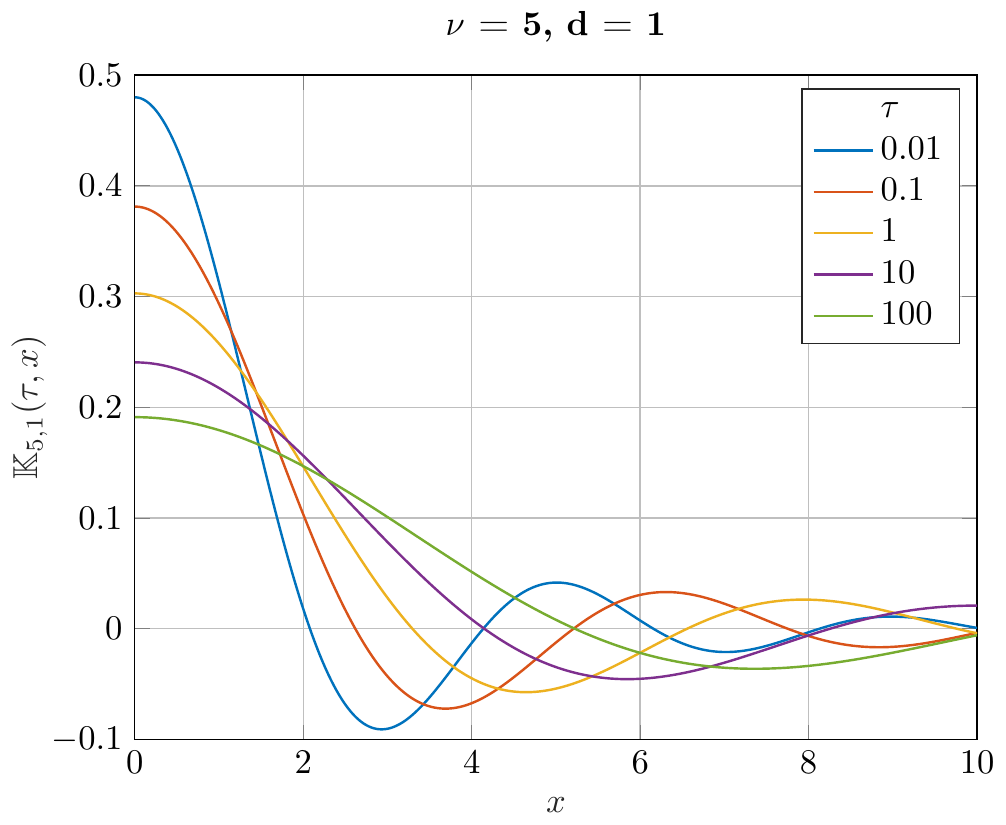}
	\caption{Graphs of the function $\bbK_{5,1}(\tau, x)$ for different values of the proper time $\tau$.} \label{Fig3} \end{center}
\end{figure}

The graphs of the functions $\calE_{\nu, \alpha}(z)$ and $\bbK_{\nu, d}(\tau, \bm{x})$ for various values of the parameters, obtained by numerical summation of the series \eqref{Em} in MATLAB, are shown in Figs. \ref{Fig1}--\ref{Fig3}. Important distinction from the case of a monotonic exponential falloff for $\nu=1$ is that the heat kernel for $\nu \ne 1$ oscillates as a function of $\bm{x}^2/\tau^{1/\nu}$.

Other interesting properties of $\calE_{\nu, \alpha}(z)$ include the following simple differentiation rule
    \begin{equation} \label{DiffRule}
    \frac{d^\beta}{dz^\beta} \calE_{\nu,\alpha}(z) = \calE_{\nu,\alpha+\beta}(z).
    \end{equation}
For integer $\beta$, it can be verified directly by differentiating the definition \eqref{EPsi}. However, this relation makes sense also for all $\beta$ such that $\calE_{\nu,\alpha+\beta}(z)$ is well defined for noninteger $\nu$ (and for any complex $\beta$ if $\nu$ is integer). For negative integer $\beta$ it will give the principal primitive of the function $\calE_{\nu, \alpha}(z)$. For noninteger $\beta$, the symbol $d^\beta/dz^\beta$ should be understood as a certain operator of fractional integrodifferentiation. Thus for each $\nu$ in the range $1/2 < \nu \le \infty$ the family of functions $\calE_{\nu, \alpha}(z)$ is closed under the operation of integrodifferentiation.

For noninteger $\nu$ the expression \eqref{FunkEvol} is the solution of the heat equation in which the nonlocal operator $(-\Delta)^\nu $ should be understood as a pseudodifferential operator defined by the Fourier transform \cite{SKM}. The corresponding equations are called fractional diffusion equations and have been widely discussed in  \cite{Pschu, ZUS, Mainardi96, Gorenflo00, Mainardi01}. However, in these papers fractional equations are usually considered in $(1+1)$-dimensional $(\tau,\bm{x})$-space , i.e. the case of $d=1$ in our notations.

\section{Integer power of Laplacian and semiclassical expansion} \label{S3}
As we see, the asymptotic behavior of GEF $\calE_{\nu,\alpha}(-z)$ at $z\to\infty$ is critically different for noninteger and integer values of $\nu$. It is power-law for noninteger $\nu$ corresponding to the nonlocal operator $(-\Delta)^\nu$ and quasiexponential $O(z^{-\infty})$ for integer $\nu$ corresponding to local differential operators of order $2\nu$. But for the heat kernel \eqref{FunkEvol} this limit is associated with the semiclassical limit $\bm{x}^2/4\tau^{1/\nu}\to\infty$, the proper time $\tau\to 0$ playing the role of $\hbar$. On the other hand, semiclassical approximation for the solution of the heat equation (or the Schr\"odinger equation in the imaginary time)
    \begin{equation} \label{EqEvol}
    \partial_\tau\mathbb{K}_F(\tau|\bm{x},\bm{y}) = -F(\nabla_x)\, \bbK_F(\tau|\bm{x},\bm{y})
    \end{equation}
begins with the Pauli--Van Vleck (or WKB) ansatz \cite{Morette}
    \begin{equation}
    \sqrt{\det\left[-\frac1{2\pi}
    \frac{\partial^2S(\tau|\bm{x},\bm{y})}{\partial x^a\,\partial y^b}\right]}\,
    \exp\left[-S(\tau|\bm{x},\bm{y})\,\right],  \label{WKB}
    \end{equation}
where $S(\tau|\bm{x},\bm{y})$ is the principal Hamilton-Jacobi function, i.e. the solution of the Hamilton-Jacobi equation with the Hamiltonian $F(\bm{p})$\footnote{The operator $F(\nabla)$ with the space-time gradients replaced by the canonical momenta $\bm{p}$. We work in Euclidean time, which explains the absence of imaginary $i$-factor. The derivation for a generic Hamiltonian not necessarily polynomial in derivatives (momenta) can be found in Appendix C of \cite{Barvinsky93}.}
    \begin{equation}
    -\frac{\partial S}{\partial\tau}
    +F\left(-\frac{\partial S}{\partial\bm{x}}\right)=0.
    \end{equation}
For the power of Laplacian this Hamiltonian equals $F(\bm{p})=(-\bm{p}^2)^\nu$, and the solution of this equation readily expresses as
    \begin{equation}
    S(\tau|\bm{x},\bm{y})=(-1)^{\frac{\nu-1}{2\nu-1}}(2\nu-1)
    \left(\frac{(\bm{x}-\bm{y})^2}
    {4\nu^2\tau^{1/\nu}}\right)^{\frac\nu{2\nu-1}}. \label{action}
    \end{equation}
This leads to the absolute value of the preexponential factor
\begin{multline}
\frac1{(2\pi)^{d/2}}\left|\det\frac{\partial^2S(\tau|\bm{x},\bm{y})}{\partial x^a\,\partial y^b}\right|^{1/2} \\
= \frac1{(4\pi\tau^{1/\nu})^{d/2}} \left(\frac{(\bm{x}-\bm{y})^2}{4\tau^{1/\nu}}\right)^{-\frac{d}2\frac{\nu-1}{2\nu-1}} \frac{\nu^{-\frac{d}2\frac1{2\nu-1}}}{\sqrt{2\nu-1}}.  \label{prefactor}
\end{multline}
Regarding its phase factor, which should be determined here by the correct choice of the branch for the fractional powers, the Pauli--Van Vleck algorithm does not give this information in contrast to the trivial case of $\nu=1$. Neither does it prescribe a definite linear combination of these branches in the heat kernel asymptotics. Below all this will be attained by two different methods --- application of the general technique of Fox $H$-functions and by the steepest descent approximation.

Note now that in view of the discussion in the previous section this quasiexponential behavior is completely impossible for noninteger $\nu$ with a power-law behavior, so that standard semiclassical expansion seems to break down for nonlocal operators $(-\Delta)^\nu$. Therefore in what follows we consider only the case of positive integer powers. To underline this, we will further denote the power of the Laplacian by $N$, when it is integer, and by $\nu$, when it can be either integer or noninteger. The goal of this section is to find the heat kernel asymptotic expansion for this higher-derivative case by an alternative method which allows to get correct complex branches and to compare them with the semiclassical result of the above type.

\subsection{Exponential asymptotics for integer order GEF} \label{S31}

Asymptotic expansion of the integral \eqref{MellinBurns} is a part of the general theory of Fox-Wright $\varPsi$- and Fox $H$-functions, briefly outlined in Appendix \ref{App1}. The idea of this expansion consists in using the Mellin transform of the gamma function
    \begin{equation} \label{InvMellin}
    \frac{1}{2\pi i} \int\limits_{w-i\infty}^{w+i\infty} ds\, \Gamma(\mu s-K) z^{-s} = \frac{1}{\mu} z^{-K/\mu}\exp\left(- z^{1/\mu}\right),
    \end{equation}
which for $\mu>0$ and growing $K>0$ will generate decreasing terms of the exponential expansion at $z\to\infty$. However, what is integrated in \eqref{MellinBurns} is not just a gamma function of this type, but rather a nontrivial ratio of those. This ratio $\varepsilon_{N,\alpha}(s)$, which is given by Eq.~\eqref{MellinCalE}, can nevertheless be converted to the series of gamma function terms of the above type, so that the $s$-integration can be successfully done. For this purpose we use, first of all, the Euler reflection formula
\begin{equation} \label{EulerRefl}
\Gamma(x)\Gamma(1-x) = \frac{\pi}{\sin(\pi x)}
\end{equation}
to provide positive coefficients of the integration parameter $s$ in the arguments of all gamma functions (just like in \eqref{InvMellin}),
\begin{gather} \label{Dop}
\varepsilon_{\nu,\alpha}(s) = \frac{1}{\nu}\frac{\sin\pi(s-\alpha)}{\sin\frac{\pi}{\nu}(s-\alpha)}\, \tilde\varepsilon_{\nu,\alpha}(s), \\
\tilde\varepsilon_{\nu,\alpha}(s) = \nu\frac{\Gamma(s)\Gamma(s-\alpha)}{\Gamma\left(\frac{s-\alpha}{\nu}\right)}.
\end{gather}
This makes the sequence of gamma function poles running to the left of the complex plane of $s$.

For integer $N$ the ratio of sines reduces to the sum of complex exponential functions
    \begin{equation}
    \frac{\sin(N\phi)}{\sin\phi} = \sum\limits_{j=0}^{N-1} e^{i(2j+1-N)\phi}, \quad \phi=\frac{\pi}{N}(s-\alpha),
    \end{equation}
After the substitution of \eqref{Dop} into the Mellin transform \eqref{MellinBurns} this leads to the sum
\begin{gather}
\calE_{N,\alpha}(-z) = \frac{1}{N} \sum\limits_{j=0}^{N-1} e^{i\omega_j\alpha}\, \tilde\calE_{N,\alpha}(-e^{i\omega_j}z), \\
\omega_j = \pi\frac{1-N+2j}{N}, \label{calEW}
\end{gather}
with the phase factors both in the coefficients of this sum and in the arguments of the new functions $\tilde\calE_{\nu,\alpha}(-e^{i\omega_j}z)$ which can be called \emph{the generalized exponential functions of the second kind}. They read
    \begin{equation}\label{1000}
    \tilde\calE_{\nu,\alpha}(-z) = \frac{\nu}{2\pi i} \int\limits_C ds\, \frac{\Gamma(s)\Gamma(s-\alpha)}{\Gamma\left(\frac{s-\alpha}{\nu}\right)} z^{-s}.
    \end{equation}

Critical point of the derivation is that now one can apply Eq.~\eqref{ProdGammaAsympt} of the Appendix \ref{App1} to expand the ratio of gamma functions in the asymptotic series of another set of gamma functions of decreasing arguments. With the parameters $p=1$, $A_1=1/\nu$, $a_1=-\alpha/\nu$, $q=2$, $B_1=B_2=1$, $b_1=0$, $b_2=-\alpha$, which give rise to the parameters \eqref{mu_beta}, \eqref{a} and \eqref{C} defined in this Appendix,
\begin{gather}\label{KappaC}
\mu = \frac{2\nu-1}{\nu}, \qquad a=\alpha\frac{\nu-1}{\nu}, \\
C = \frac{(2\nu-1)^{\alpha\frac{\nu-1}{\nu}+\frac{1}{2}}}{\nu^{1+\alpha}},
\end{gather}
Eq.~\eqref{ProdGammaAsympt} generates the asymptotic expansion
    \begin{gather}
    \tilde\varepsilon_{\nu,\alpha}(s)=
    \frac{(2\nu-1)^{a-\mu s+1/2}}{\nu^{\alpha-2s+1}}
    \sum\limits_{m=0}^\infty E_m \Gamma(\mu s-a-m), \label{1001}
    \end{gather}
with the coefficients $E_m$ independent of the integration variable $s$. These coefficients start with $E_0 = 1$ and they are systematically calculable by the procedure of Appendix \ref{App2}. The essence of this expansion is that for large $s$ it runs over ever decreasing terms, each term being smaller than the preceding one in view of the obvious relation $\Gamma(\mu s-a-m-1)=\Gamma(\mu s-a-m)/(\mu s-a-m-1)$.

The Mellin transform \eqref{1000} of \eqref{1001} on account of \eqref{InvMellin} yields the following asymptotic expansion for GEF of the second kind
\begin{multline} \label{AsymptW}
\tilde\calE_{\nu,\alpha}(-z) = \nu\frac{(\nu z^{1-\nu})^{-\frac{\alpha}{2\nu-1}}}{\sqrt{2\nu-1}} \exp\left[-(2\nu-1)\left(\frac{z}{\nu^2}\right)^{\frac{\nu}{2\nu-1}}\right] \\
\times \sum_{m=0}^\infty  \frac{E_m}{(2\nu-1)^m} \left(\frac{\nu^2}{z}\right)^{\frac{m\nu}{2\nu-1}}.
\end{multline}
In contrast to the GEF of the first kind $\calE_{\nu,\alpha}(z)$, which has no singularities, the function $\tilde\calE_{\nu,\alpha}(z)$ is singular at $z=0$. However, it always has a simple exponential (i.e., not a power-law) asymptotic behavior \eqref{AsymptW} for $z\to\infty$. Second, it is monotonic for $-\infty<z<0$ without oscillations characterizing GEF of the first kind. Moreover, one can say that the source of these oscillations is in fact the set of sines in Eq. \eqref{Dop} and phase factors in Eq.~\eqref{calEW}.

Thus in view of the decomposition \eqref{calEW} the asymptotic expansion of the generalized exponential function finally reads as a sum of $N$ series of terms
\begin{multline} \label{AsymptE2}
\calE_{N,\alpha}(-z)=\frac{N^{-\frac{\alpha}{2N-1}} z^{-\alpha\frac{N-1}{2N-1}}}{\sqrt{2N-1}} \\
\times\sum\limits_{j=0}^{N-1} \exp\left[-(2N-1)e^{i\varphi_j} \left(\frac{z}{N^2}\right)^{\frac{N}{2N-1}}+i\varphi_j\alpha\right] \\
\times\sum_{m=0}^\infty  \frac{E_m}{(2N-1)^m} \left(\frac{N^2}{z}\right)^{\frac{mN}{2N-1}} e^{-i\varphi_j m},
\end{multline}
where both the amplitudes and phases depend on the phases $\varphi_j$ of the complex factors \eqref{calEW}
    \begin{equation} \label{phase}
    \varphi_j=\frac{N}{2N-1}\omega_j=\pi\frac{1-N+2j}{2N-1},\quad j=0,1,\ldots, N-1.
    \end{equation}

Consequently, the expression in \eqref{FunkEvol} with $z=\bm{x}^2/4\tau^{1/N}$ gives the heat kernel asymptotics for fixed $\tau$ and $|\,\bm{x}\,|\to\infty$ or for fixed $|\,\bm{x}\,|$ and $\tau\to0$,
\begin{multline} \label{AsymptU}
\bbK_{N,d}(\tau,\bm{x}) = \frac1{(4\pi\tau^{1/N})^{d/2}} \left(\frac{\bm{x}^2}{4\tau^{1/N}}\right)^{-\frac{d}2\frac{N-1}{2N-1}} \frac{N^{-\frac{d}2\frac1{2N-1}}}{\sqrt{2N-1}} \\
\times\sum_{m=0}^\infty  \frac{E_m}{(2N-1)^m} \left(\frac{4N^2\tau^{1/N}}{\bm{x}^2}\right)^{\frac{mN}{2N-1}} \\
\times \sum\limits_{j=0}^{N-1}\exp\left[-(2N-1)e^{i\varphi_j} \left(\frac{\bm{x}^2}{4N^2\tau^{1/N}}\right)^{\frac{N}{2N-1}}\right] e^{i\varphi_j \left(\frac{d}{2}-m\right)}.
\end{multline}
The phase factors $e^{i\varphi_j}$ here coincide with the set of fractional powers $(-1)^{(N-1)/(2N-1)}$ in the coefficient of the Hamilton-Jacobi function \eqref{action}. This confirms the semiclassical ansatz \eqref{WKB} along with establishing a concrete choice of the linear combination of its complex branches.

The leading order term of this expansion consists of two complex conjugated branches corresponding to the maximal real part of the exponential with $j=0$ and $j=N-1$ and the phase factors $\exp(\mp i\varphi)$,
\begin{equation}
\varphi= \varphi_{N-1} = -\varphi_0=\pi \frac{N-1}{2N-1},
\end{equation}
\begin{multline} \label{AsymptU}
\bbK_{N,d}(\tau,\bm{x})\simeq\frac1{(4\pi\tau^{1/N})^{d/2}} \left(\frac{\bm{x}^2}{4\tau^{1/N}}\right)^{-\frac{d}2\frac{N-1}{2N-1}} \frac{N^{-\frac{d}2\frac1{2N-1}}}{\sqrt{2N-1}} \\
\times \exp\left[-(2N-1)e^{i\varphi} \left(\frac{\bm{x}^2}{4N^2\tau^{1/N}}\right)^{\frac{N}{2N-1}} + \frac{i\varphi d}2\right] +{\rm c.c.}
\end{multline}
and the corrections in the form of growing fractional powers of $\tau^{1/(2N-1)}\to 0$.

\subsection{Steepest descent approximation}

Alternatively this result can be reproduced by the steepest descent method which is the basis of the semiclassical approximation with the small parameter $\tau\to 0$. The change of variables $\bm{p} = \tau^{1/N} \bm{k}$, $\bm{y} = \tau^{(N-1)/N} \bm{x}$, converts the momentum integral \eqref{UIntP} to the form
\begin{gather} \label{StPhU}
\bbK_{N,d}(\tau,\bm{x}) = \frac1{(2\pi\tau^{1/N})^d} \int d^dp\, e^{-S(\bm{p})/\tau}, \\
\text{where}\quad S(\bm{p}) = (\bm{p}^2)^N - i\bm{p}\bm{y}.
\end{gather}
Its short time $\tau\to 0$ asymptotics follows from the saddle points of the complex action $S(\bm{p})$ at which this action is stationary, $\partial S(\bm{p})/\partial p_a=0$. These $2N-1$ points read
    \begin{gather}
    \bm{p}_j = \left[\,i^{\frac1{2N-1}}\,\right]_j
    \left(\frac{\tau^\frac{N-1}{N}|\,\bm{x}\,|}{2N}\right)^{\frac{1}{2N-1}}\,
    \frac{\bm{x}}{|\,\bm{x}\,|},
    \end{gather}
where $\left[\,i^{\frac1{2N-1}}\,\right]_j$, $j=0,1,..., 2N-2$, are the $2N-1$ roots of the imaginary unit $i$. The leading order contribution of each such saddle point is given by the standard expression
\begin{equation} \label{Pereval}
\bbK_{N,d}^{(j)}(\tau,\bm{x}) = \left. \frac{e^{-S(\bm{p})/\tau}}{(2\pi\tau^{1/N})^d} \left(\det\frac1{2\pi\tau}\frac{\partial^2S(\bm{p})}{\partial p_a\partial p_b}\right)^{-1/2} \,\right|_{\,\bm{p}=\bm{p}_j},
\end{equation}
which gives
\begin{multline} \label{AsymptU1}
\bbK^{(j)}_{N,d}(\tau,\bm{x})=\frac1{(4\pi\tau^{1/N})^{d/2}} \left(\frac{\bm{x}^2}{4\tau^{1/N}}\right)^{-\frac{d}2\frac{N-1}{2N-1}} \frac{N^{-\frac{d}2\frac1{2N-1}}}{\sqrt{2N-1}} \\
\times \exp\left[-(2N-1)e^{i\varphi_j} \left(\frac{\bm{x}^2}{4N^2\tau^{1/N}}\right)^{\frac{N}{2N-1}} + \frac{i\varphi_j d}2\right],
\end{multline}
where the set of phase factors  is determined by the relations
\begin{equation} \label{phase1}
e^{i\varphi_j} = -i \left[\,i^{\frac1{2N-1}}\,\right]_j, \qquad \varphi_j=\pi\frac{1-N+2j}{2N-1},
\end{equation}
$j=0,1,\ldots, 2N-2,$ and in fact extends the range \eqref{phase} up to $2N-1$.

The most complicated part of the saddle point method is the proof of the existence of a correct steepest descent integration contour and the choice of relevant saddle points which should contribute to the asymptotics in question \cite{Fedoryuk}. In a particular case of the integral \eqref{UIntP} there is an obvious hint on their choice, confirmed by a rigorous analysis in \cite{Fedoryuk}, that the points with $\cos\varphi_j<0$, $j\ge N$, contribute the terms exponentially growing with $|\,\bm{x}\,|$, which contradicts the known $|\,\bm{x}\,|^{-\infty}$ falloff at $|\,\bm{x}\,|\to\infty$. Therefore only the remaining $N$ points with $j\le N-1$ can contribute to the asymptotics of the integral. Note that their respective phase factors coincide with those defined by the Eqs.~\eqref{calEW} and \eqref{phase}, and thus correspond to the set of terms in the decomposition of the generalized exponential function into the sum of GEF of the second kind. Thus the steepest descent method leads to the same result as the Mellin transform within the formalism of Fox $H$-functions.

\subsection{Nonuniformity of the semiclassical expansion} \label{S33}

Semiclassical expansion \eqref{AsymptU} clearly shows a principal difference of higher-order operators from the second-order case $N=1$. For $N>1$ the short time expansion \eqref{AsymptU}, $\tau\to 0$, does not stand the limit $|\,\bm{x}\,|\to 0$ because it involves negative powers of $|\,\bm{x}\,|$. In particular, it does not maintain the initial condition $\bbK_{\nu,d}(0,\bm{x})=\delta(\bm{x})$. The exact heat kernel of course satisfies this condition, because for any smooth test function $f(\bm{x})$
\begin{multline}
\int d^dx\, \bbK_{\nu,d}(\tau,\bm{x}) f(\bm{x}) \\
= \frac{1}{(4\pi)^{d/2}} \int d^dy\, \calE_{\nu,\frac{d}{2}}\left(-\frac{\bm{y}^2}{4}\right) f(\tau^{1/\nu}\bm{y}) \xrightarrow[\tau\to 0]{} f(0),
\end{multline}
since
\begin{multline}
\frac1{(4\pi)^{d/2}} \int d^dy\, \calE_{\nu,\alpha}\left(-\frac{\bm{y}^2}{4}\right) \\
= \frac1{\Gamma(d/2)} \int\limits_0^\infty dz\,z^{\frac{d}{2}-1} \calE_{\nu,\alpha}(-z) = \frac{\Gamma\left(\frac{\alpha-d/2}{\nu}\right)}{\nu\Gamma\left(\alpha-\frac{d}{2}\right)} \xrightarrow[\alpha\to\frac{d}{2}]{}1,
\end{multline}
where we used the relation \eqref{MellinCalE}. Interestingly, in the recovery of this result for integer $N$ each of the $N$ terms of the decomposition \eqref{calEW} contributes one and the same $1/N$th part of it, because their dependence on $\omega_j$ drops out in the limit $\alpha\to d/2$,
\begin{multline}
\frac1{(4\pi)^{d/2}} \int d^dy\, \frac{e^{i\alpha\omega_j}}{N} \tilde\calE_{\nu,\alpha}\left(-e^{i\omega_j}\frac{\bm{y}^2}{4}\right) \\
=  \frac{\Gamma\left(\alpha-\frac{d}{2}\right)}{\Gamma\left(\frac{\alpha-d/2}{N}\right)} e^{i\omega_j\left(\alpha-\frac{d}{2}\right)}
\xrightarrow[\alpha\to d/2]{}\frac{1}{N}.        \label{integral}
\end{multline}

At the same time, if we try to reproduce the same result by using $N$ branches \eqref{AsymptU1} of the semiclassical expansion, corresponding to different terms of the above decomposition
    \begin{eqnarray} \label{Asympt3}
    \bbK_{N,d}(\tau,\bm{x})=\sum\limits_{j=0}^{N-1}
    \bbK^{(j)}_{N,d}(\tau,\bm{x})\,
    \left[\,1+ O\left(\tau^{\frac1{2N-1}}\right)\,\right],
    \end{eqnarray}
then the result will be critically different. Each $\bbK^{(j)}_{N,d}(\tau,\bm{x})$ is singular at $\bm{x}=0$, but this singularity is integrable. However, the result of this integration is different from \eqref{integral}
    \begin{eqnarray} \label{Asympt3}
    \int d^dx\,\bbK^{(j)}_{N,d}(\tau,\bm{x})
    =\frac{N^{\frac{d}2-1}}{(2N-1)^{\frac{d-1}2}}.
    \end{eqnarray}

Even more striking discrepancy between the exact heat kernel and its asymptotics is that, while all the terms of the latter are singular in the limit $\bm{x}\to 0$, the GEF \eqref{Em} and the exact $\bbK_{\nu,d}(\tau,\bm{x})$ are both well defined in this limit
    \begin{eqnarray} \label{zero_limit}
    \bbK_{\nu,d}(\tau,0)
    =\frac1{\big(4\pi\tau^{1/\nu}\big)^{d/2}}\,
    \frac{\Gamma(d/2\nu)}{\nu
    \Gamma(d/2)}.
    \end{eqnarray}
The short time expansion of this coincidence limit which is a main goal of the Seeley-Gilkey technique \cite{Seeley, Gilkey1975, Gilkey1979, Gilkey2003} runs in powers of $\tau^{1/\nu}$, whereas the expansion \eqref{AsymptU} goes in the other fractional powers  $\tau^{1/(2N-1)}$.

The source of all these discrepancies\footnote{There is additional controversy with the result \eqref{Asympt3}---while all $\bbK^{(j)}_{N,d}(\tau,\bm{x})$ with $j\neq 0$ and $j\neq N-1$ are exponentially subdominant and should be discarded according to asymptotic expansion theory, all their integrals \eqref{Asympt3} are of the same order of magnitude.} is the fact that, contrary to the second order Laplacian, the heat kernel asymptotic expansion is not uniform in $\bm{x}\to 0$. While in this limit the expansion \eqref{HeatKernel0} for a minimal second order operator $F(\nabla)=-\Delta+\ldots$ is just a regularization of the delta function, for $N>1$ the expansion \eqref{AsymptU} fails for $\bm{x}\to 0$ and does not have a chance to reproduce correct initial conditions with a pointlike support at $\bm{x}=0$. Obviously, there is no such a discrepancy in the case of $N=1$ with a single $j=0$ branch of the heat kernel expansion, so that the coincidence limit $\bm{y}=\bm{x}$ can be directly taken in the asymptotic expansion \eqref{HeatKernel0}.

\section{Conclusions}

Thus we obtained the expression \eqref{FunkEvol} for the heat kernel $\bbK_{\nu, d}(\tau, \bm{x})$ of the operator $(-\Delta)^\nu$ in the $d$-dimensional flat space, which is a direct generalization of the well-known heat kernel \eqref{HeatKernel} to local higher derivative and nonlocal (pseudodifferential) operators. This generalization is represented in terms of the newly introduced two-parameter family of generalized exponential functions $\calE_{\nu,\alpha}(z)$, $\alpha=d/2$, $z=-\bm{x}^2/\tau^{1/\nu}$, defined by the Taylor series \eqref{Em} and related to Fox-Wright $\varPsi$- and Fox $H$-functions. We studied various properties of these functions and their integral representations. They include, in particular, the Mellin-Barnes representation which allows one to find, by the technique of Fox $H$-functions, their asymptotic expansion in the limit of $z\to\infty$ corresponding to two equivalent asymptotics of the heat kernel as $\tau\to 0$ or as $|\bm{x}|\to\infty$.

This expansion turns out to be critically different for integer and noninteger values of $\nu$. In contrast to the exponential behavior, anticipated on the ground of semiclassical considerations with $\tau\to 0$ playing the role of $\hbar$, for noninteger $\nu$, that is for nonlocal operators  $(-\Delta)^\nu$, this is a power-law falloff. For integer $\nu>1$ this asymptotic expansion matches with the exponential Pauli--Van Vleck ansatz or steepest descent approximation, but it essentially differs from the pure Laplacian case of $\nu=1$. In particular, this asymptotic expansion is not uniform for all values of $|\,\bm{x}\,|\to 0$ and does not stand a singular coincidence limit $\bm{x}=0$ which is, on the other hand, easily accessible directly from the short time expansion for $\nu=1$. In addition, the heat kernel for higher-derivative and nonlocal operators with $\nu\ne1$ features oscillatory behavior in $\bm{x}$-space contrary to the monotonic exponential falloff for the pure Laplacian.

The coincidence limit of the heat kernel $\bbK_F(\tau|\bm{x},\bm{x})$ and its functional trace
    \begin{equation}
    \Tr e^{-\tau F} = \int d^dx\,\tr\bbK_F(\tau|\bm{x},\bm{x})  \label{5.1}
    \end{equation}
are the objects of major interest in the Schwinger--DeWitt technique for quantum effective action. For generic minimal second order operators with $\bm{x}$-dependent coefficients it is based on the expansion \eqref{HeatKernel0} for $\bbK_F(\tau|\bm{x},\bm{y})$ with split arguments (generically lacking translation invariance).  This coincidence limit is also the subject of mathematical Seeley--Gilkey treatise by various functorial methods not directly appealing to off-diagonal elements of $\bbK_F(\tau|\bm{x},\bm{y})$. The absence of a nonsingular coincidence limit in the heat kernel asymptotic expansion for $\nu\ne1$ considered above seems to invalidate any attempt to use it for some generalization of the Schwinger--DeWitt method. So these expansions are not very physically interesting in field theory applications. However, a systematic way of the short time expansion of the heat kernel, which underlies UV properties of field models, is provided by the recurrent equations for the coefficients of the expansion \eqref{HeatKernel0}, these equations heavily relying on the off-diagonal $\bbK_F(\tau|\bm{x},\bm{y})$. The Seeley--Gilkey functorial methods are not so universal and powerful enough to yield everything what physicists need in quantum gravity and other applications. For example, Ho\v{r}ava--Lifshitz gravity \cite{Barvinsky17, Barvinsky172} are encumbered with the necessity of working with higher order and nonminimal operators whose leading symbol does not reduce to the power of Laplacian and, therefore, go outside of the scope of functorial methods. Derivation of recurrent equations for the two-point coefficients of \eqref{HeatKernel0} generalized to such operators then becomes indispensable.

The generalized exponential functions introduced above provide fundamental building blocks of such recurrent equations, and the lack of uniformity of their asymptotics does not make them less efficient. Note that the characteristic feature of the Schwinger--DeWitt expansion \eqref{HeatKernel0} is a single overall exponential factor absorbing all essentially singular dependence on $\tau\to 0$. The attempt to directly generalize this expansion to $\nu\ne1$ with a single semiclassical exponential factor fails because it generates infinitely many negative powers of $\tau$.

On the contrary, resummation of these singular terms can be performed with the aid of the generalized exponential functions, but in contrast to \eqref{HeatKernel0} these functions will not form a single overall factor, but rather comprise the series of terms with different $\alpha$-parameters. As we are going to show in the coming paper \cite{Wach2}, for a minimal \emph{differential} operator $F$ of an (integer) order $2N$ in a curved Riemannian space [$(\bm{x}-\bm{y})^2/2\to\sigma(\bm{x},\bm{y})$] the following generalization of the expansion \eqref{HeatKernel0} holds
\begin{multline} \label{HeatKernel5}
\bbK_F(\tau|\bm{x},\bm{y}) = \frac1{(4\pi\tau^{1/N})^{d/2}} \\
\times\sum\limits_{j=0}^\infty \tau^{j/N} \calE_{N,\frac{d}{2}-j}\left(-\frac{\sigma(\bm{x},\bm{y})}{2\tau^{1/N}}\right) a_j[F|\bm{x},\bm{y}].
\end{multline}
The generalized HaMiDeW-coefficients $a_j[F|\bm{x},\bm{y}]$ satisfy the manageable chain of recurrent equations, which can be solved for the coincidence limit $\bm{x} = \bm{y}$. Note also that the coincidence limit of \eqref{HeatKernel5} is well defined, even though the asymptotics of the underlying $\calE_{N,\alpha}(-z)$ are not uniform for $z\to 0$. In fact, these asymptotics are not needed for this limit. Since $\calE_{N,\alpha}(0) = \Gamma(\alpha/N)/N\Gamma(\alpha)$ we have the following expansion for the coincidence limit
    \begin{eqnarray}
    \bbK_F(\tau|\bm{x},\bm{x}) = \tau^{-d/2N} \sum\limits_{j=0}^\infty \tau^{j/N} A_j[F|\bm{x}], \label{Kxx}
    \end{eqnarray}
where
    \begin{eqnarray}
    A_j[F|\bm{x}] = \frac{1}{(4\pi)^{d/2}} \frac{\Gamma\left(\frac{d/2-j}{N}\right)}{N\Gamma(d/2-j)} a_j[F|\bm{x},\bm{x}]. \label{Aa}
    \end{eqnarray}

So GEF should be treated as entire building blocks of the formalism, the operations with them being based on their simple differentiation rule \eqref{DiffRule} and the value at $z=0$. In our next paper \cite{Wach2} we will consider various properties of the generalized HaMiDeW-coefficients in \eqref{HeatKernel5}. In particular, we will prove the generalized ``functorial property'' for an arbitrary power $\lambda$ of a differential operator $F$,
\begin{equation}
a_j[F^\lambda|\bm{x},\bm{y}] = a_j[F|\bm{x},\bm{y}]
\end{equation}
(previously known only in the coincidence limit  $\bm{x}=\bm{y}$ \cite{Gilkey1980}), and also easily reproduce and extend the results of \cite{Gilkey1980} for higher order operators. These coefficients and the computational methods based on them promise to be very efficient and are likely to simplify the calculation of beta functions for theories with higher derivatives and Ho\v{r}ava--Lifshitz type models \cite{Barvinsky17,Barvinsky172}. All this makes the use of GEF and the associated heat kernel coefficients very prospective.

The above formalism seems equally applicable to the case of generic noninteger $\nu$. This case is, in particular, important in superrenormalizable quantum gravity models \cite{Tomb, Modesto, Talaganis, Biswas}, within analytical regularization of Feynman graphs \cite{Tarasov18} or for the calculation of UV counterterms in Ho\v{r}ava--Lifshitz gravity models. For example, in (3+1)-dimensional Ho\v{r}ava gravity cubic in spatial curvature counterterms follow from the heat kernel of the operator which is a square root of the sixth order nonminimal differential operator \cite{Barvinsky172}. However, as we saw above there is a number of new features (like the power-law heat kernel asymptotics confronting their exponential analogue for integer $N$ or the mismatch with the semiclassical expansion) which might backfire under indiscriminate extension of this method. Here we only briefly comment on possible modifications due to these subtleties.

One modification follows from the recovery of the heat kernel diagonal elements by inverse Mellin transform from the operator zeta function $F^{-s}\delta(\bm{x},\bm{y})\,|_{\bm{x}=\bm{y}}$ used in \cite{Bar2003}. For the operators of the form $F = H^\nu$, where $H$ is a Laplace type (minimal second order) operator, and noninteger $\nu$ this method leads to additional terms \cite{Bar2003}
\begin{equation}
	\bbK_{F}(\tau|\bm{x},\bm{x}) = \tau^{-d/2\nu} \sum\limits_{j=0}^\infty \tau^{j/\nu} A_j(\bm{x})
	+ \sum\limits_{k=1}^\infty \tau^k B_k(\bm{x}).  \label{5.5}
\end{equation}
While the coefficients $A_j(\bm{x})$ are in one-to-one correspondence with the HaMiDeW-coefficients $a_j[H|\bm{x},\bm{x}]$ of the operator $H$ and are local quantities, the coefficients $B_k(\bm{x})$ are determined through the values of the zeta function at certain values of $s$. Rather than being expressed in terms of $a_j[F|\bm{x},\bm{x}]$, they turn out to be nonlocal and irrelevant to UV renormalization because they do not contribute to UV divergences in view of analytic expansion in $\tau$ starting with a linear term. However, according to our method, just in the case of operators of the form $F = H^\nu$, such additional terms do not arise.

Another modification known from mathematical studies \cite{GilkeyGrubb, Bar2003} is the origin of logarithmic terms in the proper time expansion of the heat kernel. For special values of noninteger $\nu$ leading to $-\alpha=n+k\nu \neq m$ [with positive integer $k$, $n$ and $m$---see the discussion after \eqref{UslOpr}] the functions $\calE_{\nu,\alpha}(z)$ in the expansion \eqref{HeatKernel} are not defined because of gamma function singularities. This exceptional case can occur, in particular, for even order roots of the Laplace type operator in odd spacetime dimensions. In this case the logarithmic terms appear \cite{GilkeyGrubb, Bar2003}
\begin{equation}
	\bbK_{F}(\tau|\bm{x},\bm{x}) = \tau^{-d/2\nu} \sum\limits_{j=0}^\infty \tau^{j/\nu} A_j(\bm{x})
	+ \sum\limits_{k=1}^\infty \tau^k \log\tau\, C_k(\bm{x}).
\end{equation}
which are again unrelated to renormalization of UV divergences.

All these modifications can apparently be attributed to the fact that zeta-function approach of \cite{GilkeyGrubb, Bar2003} actually represents a regularization which for nonlocal theories (corresponding to noninteger values of $\nu$) leads to different results\footnote{Note that the method of derivation of \eqref{5.5} in \cite{Bar2003} can be interpreted as zeta-function regularization, because it operates with the regularized (and therefore finite) expression for the coincidence limit of the Green's function of the operator $F^s$. On the contrary, our expansion is done for separate arguments of the heat kernel which of course renders  this coincidence limit singular and invokes point separation or dimensional regularization.}. Absence of uniformity of the asymptotic small time expansion in the vicinity of the heat kernel diagonal discussed above shows up for noninteger values of $\nu$. The search for an asymptotic expansion of GEF and heat kernel that would be uniform for all $\bm{x}$ and $\bm{y}$ (analogous, for example, to the uniform WKB asymptotic expansion of Legendre functions \cite{Thorne}) apparently could have resolved the problem of these discrepancies. This however goes beyond the scope of this paper, in particular, because nonuniformity of the asymptotic expansion of GEF is harmless in the calculation of the functional trace \eqref{5.1} if one uses generalized exponential functions as building blocks of the expansion and takes their exact values at zero argument.

To summarize, the generalized exponential functions can serve as a very efficient tool in quantum field theory and quantum gravity. Moreover, their connection with fractional calculus opens the prospect of applying the obtained heat kernels far beyond the area of QFT. This includes the theory of fractional differential equations which can be effectively used to construct phenomenological models of fractal media, systems with memory and nonlocal interaction. Numerous applications of fractional calculus to physical problems are discussed, for example, in \cite{Tarasov} and references therein.

\section*{Acknowledgements}

The authors are grateful to A.\,E.~Kazantsev, A.\,A.~Lobashev and M.\,M.~Popova for numerous fruitful discussions, and especially to O.\,I.~Marichev for the help with the asymptotic expansion of the Fox--Wright $\varPsi$-functions. We also thank the anonymous referee for constructive suggestions on the extension of our results. This work was supported by the RFBR Grant No.17-02-00651 and by the Foundation for the Advancement of Theoretical Physics and Mathematics ``BASIS.''

\appendix
\section{Wright $\varPsi$-functions and Fox--Wright $H$-functions} \label{App1}

The Fox--Wright $\varPsi$-functions ${}_p\Psi_q[(a,A); (b,B); z]$ are labeled by two sets of parameters $A_k, a_k$, $r=1,\ldots,p$, and $B_j, b_j$, $j=1,\ldots,q$, among which $A_k$ and $B_j$ are real and positive. These functions are defined by their Taylor series
    \begin{equation} \label{PsiFunction}
    {}_p\Psi_q[(a,A); (b,B); z] = \sum\limits_{k=0}^\infty \frac{\prod\limits_{j=1}^p \Gamma(a_j+A_jk)}{\prod\limits_{i=1}^q \Gamma(b_i+B_ik)} \frac{z^k}{k!}\, .
    \end{equation}
They represent one of the possible further extensions of the generalized hypergeometric series, ${}_pF_q[a; b; z] = {}_p\Psi_q[(a,1); (b,1); z]\Gamma(b)/\Gamma(a)$, and have applications, in particular, in fractional calculus \cite{SKM, Pschu, Mainardi96, Gorenflo00, Mainardi01, Kilbas02, Kilbas05, Lav17}. They were introduced by E.~M.~Wright, who studied their asymptotic behavior \cite{Wright35, Wright40}.

In their turn the Fox--Wright $\varPsi$-functions form a special case of more general Fox $H$-functions $H_{p,q}^{m,n}\left[z \big|\begin{smallmatrix}(a,A)\\ (b, B)\end{smallmatrix}\right]$.
They are defined by the Mellin--Barnes integral
\begin{gather}
H_{p,q}^{m,n}\left[z \left|\begin{smallmatrix}(a,A)\\ (b, B)\end{smallmatrix}\right.\right] = \frac{1}{2\pi i} \int\limits_C h_{p,q}^{m,n}[s] z^{-s} ds, \label{DefHfunction} \\
h_{p,q}^{m,n}[s] = \frac{\prod\limits_{i=1}^m \Gamma(b_i+B_is) \prod\limits_{j=1}^n \Gamma(1-a_j-A_js)}{\prod\limits_{i=m+1}^q \Gamma(1-b_i-B_is) \prod\limits_{j=n+1}^p \Gamma(a_j+A_js)}, \label{DefHfunction2}
\end{gather}
also with real and positive $A_i$ and $B_j$. The poles $l_{i,k}$ of the gamma functions $\Gamma(b_i+B_is)$, $i=1,\ldots, m$, enumerating the poles index $k$ being integer, run to the left of the complex plane of $s$, whereas the poles $r_{j,k}$ of $\Gamma(1-a_j-A_js)$, $j=1,\ldots, n$, run to the right. It is assumed that the parameters $A_j$, $a_j$, $B_i$ and $b_i$ are such that these poles do not match, $l_{i,k} \ne r_{j,l}$. Then the contour of integration $C$ is chosen to pass from $-i\infty$ to $i\infty$ and to separate the poles $l_{i,k}$ and $r_{j,k}$.

The Fox $H$-functions are related to Fox--Wright $\varPsi$-functions in exactly the same way as the well-known Meyer $G$-functions to generalized hypergeometric functions. Obviously
    \begin{equation} \label{PsiViaH}
    {}_p\Psi_q\left[\left.\begin{smallmatrix}(a,A)\\ (b, B)\end{smallmatrix}\right| z\right] = H_{p,q+1}^{1,p}\left[-z \left|\begin{smallmatrix}(1-a,A)\\ (0,1), (1-b, B)\end{smallmatrix}\right.\right].
    \end{equation}

The general theory of $H$-functions and $H$-transforms can be found in \cite{Braaksma, Marichev, Sri84, MSH, KS}. Here we briefly sketch their main properties and the way of handling their asymptotic behavior. This behavior is characterized by the following three combinations of their parameters
\begin{gather}
\mu = \sum_{j=1}^{q} B_j - \sum_{k=1}^{p} A_k, \qquad \beta = \frac{\prod\limits_{k=1}^{p}A_k^{A_k}}{\prod\limits_{j=1}^{q}B_j^{B_j}}, \label{mu_beta}\\
a = \sum_{k=1}^{p} a_k - \sum_{j=1}^{q} b_j + \frac{1}{2}(q-p-1), \label{a}
\end{gather}
Note that the structure of the expression \eqref{DefHfunction2} allows one to relocate gamma functions between the numerator and the denominator using the Euler reflection formula \eqref{EulerRefl}. Under this operation only the parameters $m$ and $n$ change, while the parameters $p$, $q$, $\mu$, $\beta$ and $a$, as it is easy to see, remain intact.

The main result, based on the use of the Stirling formula, is as follows: for $\mu>0$ the contour $C$ in \eqref{DefHfunction} can be closed on the left of the complex plane, then $m$ series obtained by summing the residues at the poles $l_{i,k}$ will converge absolutely on the whole complex $z$ plane, defining, generally speaking, a multivalued function with an essentially singular point at $z=\infty$. If in this case we formally close the contour $C$ on the right, then the sum of the residues at the poles $r_{j,k}$ will determine the asymptotic (divergent) power series as $z\to\infty$. For $\mu<0$ the situation will be exactly the opposite: the sum of residues at the poles $r_{j,k}$ will absolutely converge at $z\ne0$, and the divergent series of residues at the poles $l_{i,k}$ will determine the asymptotic behavior of the function at $z\to0$. Finally, in the case of the critical value $\mu=0$ the series obtained by closing the contour $C$ on the left will converge inside the circle $|z|<\beta^{-1}$, and the series obtained by closing the contour $C$ on the right will converge outside of it.

Exponential asymptotic behavior for $z\to\infty$ appears when $h_{p,q}^{m,n}[s]$ does not have any rightgoing poles $r_{j,k}$, i.e., when in the expression \eqref{DefHfunction2} either the functions $\Gamma(1-a_j-A_js)$ in the numerator are completely absent ($n=0$), or all their poles are canceled with the poles of the functions $\Gamma(1-b_i-B_is)$ in the denominator (as it happens in the case of the functions $\calE_{\nu,\alpha}(z)$ for an integer $\nu$). The general recipe for finding exponential asymptotics, which is explained in detail in \cite{Braaksma}, is the following: one first needs to use the Euler reflection formula to relocate gamma functions so that they have only leftgoing poles, i.e. to convert the expression \eqref{DefHfunction2} to the form, when all gamma functions with the coefficients $A_j$ are in the denominator, and all gamma functions with the coefficients $B_i$ are in the numerator. After that, one needs to use the asymptotic expansion for the ratio of gamma function products,
    \begin{equation} \label{ProdGammaAsympt}
    \frac{\prod\limits_{j=1}^q \Gamma(B_js + b_j)}{\prod\limits_{k=1}^p \Gamma(A_ks + a_k)}=C (\beta\mu^\mu)^{-s} \sum\limits_{m=0}^\infty E_m\Gamma(\mu s - a - m),
    \end{equation}
which is derived by the method sketched in Appendix \ref{App2}. Here $|s|\to\infty$, $|\pi-\arg s|>\epsilon$, the parameters $\mu$, $\beta$ and $a$ are defined as above in \eqref{mu_beta} and \eqref{a},
    \begin{equation}
    C = (2\pi)^{(q-p-1)/2} \mu^{a+1/2}
    \prod\limits_{k=1}^{p} A_k^{1/2 - a_k}
    \prod\limits_{j=1}^{q} B_j^{b_j-1/2},     \label{C}
    \end{equation}
$E_0 = 1$ and other coefficients $E_m$ are systematically calculable by the method also sketched below in Appendix \ref{App2}. Finally, application of the inverse Mellin transform \eqref{InvMellin} yields the required asymptotic expansion. This procedure is used in Sec.~\ref{S3} for the derivation of the large $z$ expansion of the generalized exponential function and the associated heat kernel.

\section{The ratio of gamma function products} \label{App2}

Here we briefly sketch the details of a special asymptotic expansion at $s\to\infty$ of the ratio of two products of gamma functions in Eq.\eqref{ProdGammaAsympt}. If we denote this ratio by $R(s)$ and divide it by $\Gamma(\mu s-a)$, then in view of the Stirling expansion
\begin{multline} \label{Stirling}
\Gamma(s+x) = \sqrt{2\pi}\,e^{-s}\,s^{s+x-1/2} \\
\times\exp\left[\,\sum_{k=1}^\infty \frac{(-1)^{k+1}}{k(k+1)}B_{k+1}(x) s^{-k}\right],
\end{multline}
($B_k(x)$ are Bernoulli polynomials) the result will read as
    \begin{eqnarray}
    &&\frac{R(s)}{\Gamma(\mu s-a)} = C\,(\beta\mu^\mu)^{-s}\,
    \exp\left[\;\sum_{n=1}^\infty D_n s^{-n}\right], \label{Asympt1}
    \end{eqnarray}
where the parameters $\mu$, $\beta$, $a$, and $C$ are defined by Eqs.~\eqref{mu_beta}, \eqref{a} and \eqref{C} and the coefficients $D_n$ equal
\begin{multline}
D_n = \frac{(-1)^{n+1}}{n(n+1)} \left(\sum\limits_{j=1}^q \frac{B_{n+1}(b_j)}{B_j^n} \right. \\
 - \left.\sum\limits_{k=1}^p \frac{B_{n+1}(a_k)}{A_k^n} - \frac{B_{n+1}(-a)}{\mu^n}\right).
\end{multline}
The factor of gamma function $\Gamma(\mu s - a)$ was especially added in the left-hand side of \eqref{Asympt1} in order to cancel the powers of $s$ and $s^s$.

Now the exponential factor in \eqref{Asympt1} can be reexpanded in inverse powers of $s$ to give
    \begin{equation} \label{ExpExp}
    \exp\left[\sum_{n=1}^\infty D_n s^{-n}\right] = 1 + \sum_{n=1}^\infty C_n s^{-n},
    \end{equation}
where each coefficient $C_k$ is uniquely determined by the first $k$ coefficients $D_1, \ldots, D_k$, $C_1=D_1$, $C_2 = D_2 + D_1^2/2!$, $C_3 = D_3 + D_1D_2 + D_1^3/3!$, etc.

The next step is to rearrange the $1/s$-expansion here in terms of the inverse Pochhammer symbols for a special choice of the argument $x=a+1-\mu s$ composed of the parameters $s$, $\mu$ and $a$,
    \begin{equation}
    \frac{\Gamma(x)}{\Gamma(x+k)} = \prod_{n=1}^k \frac1{x+n-1}.
    \end{equation}
To begin with, this symbol can be expanded in inverse powers of $s$,
\begin{multline} \label{Direct}
\frac{\Gamma(a+1-\mu s)}{\Gamma(a+1-\mu s+k)}= (-1)^k\frac{\Gamma(\mu s-a-k)}{\Gamma(\mu s-a)} \\
= (-1)^k\sum_{j=1}^\infty d_{kj} s^{-j},
\end{multline}
where the coefficients of the infinite lower-triangular matrix $[d_{kj}]$ depend on $\mu$ and $a$, $d_{kj}=0$ for $j<k$, $d_{kk}=\mu^{-k}$. Inversion of this relation allows one to expand $s^{-k}$ in terms of the sequence of such symbols
    \begin{gather} \label{Inverse}
    s^{-k} = \sum_{j=1}^\infty  d_{kj}^{-1}\,
    \frac{\Gamma(\mu s-a-j)}{\Gamma(\mu s-a)},
    \end{gather}
where $d_{kj}^{-1}$ are the coefficients of the inverse matrix, $d_{kj}^{-1}=0$ for $j<k$, $d_{kk}^{-1} = \mu^k$. Using the relations \eqref{Direct} and \eqref{Inverse} we can trade the expansion in powers of $1/s$ for the expansion in $\Gamma(\mu s-a-j)/\Gamma(\mu s-a)$,
    \begin{gather}
    \sum_{k=1}^\infty C_k s^{-k} = \sum_{j=1}^\infty E_j\frac{\Gamma(\mu s-a-j)}{\Gamma(\mu s-a)}, \quad E_j = \sum\limits_{k=1}^\infty C_k d_{kj}^{-1}. \label{Pochg2}
    \end{gather}

Then using \eqref{ExpExp} and \eqref{Pochg2} in \eqref{Asympt1} and multiplying the result by $\Gamma(\mu s - a)$ we finally get the expansion \eqref{ProdGammaAsympt}.

\bibliography{HeatKernel}

\end{document}